\documentclass{elsart}
\usepackage{graphicx}
\usepackage{amsmath}
\usepackage{amssymb}

\begin{document}


\begin{frontmatter}

\title{Chalker-Coddington model described by an S-matrix with odd dimensions}

\author[address1]{Kosuke~Hirose\thanksref{thank1}},
\author[address1]{Tomi~Ohtsuki}
and
\author[address2]{Keith~Slevin}

\address[address1]{Department of Physics, Sophia University,
Tokyo 102-8554, Japan}

\address[address2]{Department of Physics, Osaka University,
Osaka 560-0043, Japan}

\thanks[thank1]{
Corresponding author.
E-mail: kousuk-h@sophia.ac.jp,
Fax:81332383341}
\begin{abstract}
The Chalker-Coddington network model is often used to describe the transport
properties of quantum Hall systems.
By adding an extra channel to this model, we introduce an asymmetric model with profoundly
different transport properties.
We present a numerical analysis of these transport properties and consider
the relevance for realistic systems.
\end{abstract}

\end{frontmatter}

\section{Introduction}

The Chalker-Coddington network model (C-C model) has been accepted
as one of the simplest models \cite{C-C,Kramer} to explain and analyze
the quantum Hall effect \cite{Klitzing}.
In the C-C model, the equipotential lines of the long-ranged disordered potential
are regarded as the links of a network,
while the saddle points of these equipotential lines are
regarded as the nodes of this network.
Quantum tunneling at the nodes of the network is described by a $2 \times 2$ scattering matrix \textbf{S}$_2$
\begin{equation}
\textbf{S}_2=\left(
\begin{array}{cc}
-r&t\\
t&r
\end{array}
\right)\,,
\end{equation}
where $r$ and $t$ are reflection and transmission amplitude, respectively.
They are real numbers and satisfy $r^2+t^2=1$.

In this paper we introduce the asymmetric Chalker-Coddington network model.
This is obtained from the standard C-C model by adding an extra channel
in such a way that the number of incoming fluxes on one side of the system
differs from that on the other side (see Fig. \ref{NovelModel}).
This leads to the creation of a single perfectly transmitting channel.
We analyze the conductance distribution for the square geometry (2D), and
the size dependence of the average conductance in both 2D and quasi-one dimensional
(quasi-1D) geometries.
The results are interpreted using random matrix theory and the repulsion of transmission
eigenvalues from the perfectly transmitting channel.
We emphasize that the asymmetry due to the addition of single channel dramatically changes the
transport properties.
This asymmetry can be realized in parallel Hall bars where the contact does not touch one edge of
a bar.
Another possibility is the transport in graphene,
where $N$ ($N'$) left (right) going and  $N'$  ($N$) right (left) going channels
are realized near $K$ ($K'$) point.  For long ranged impurities, scattering between states near
$K$ and $K'$ is supressed, resulting in the asymmetic number of left and right going channels
\cite{Sigrist}.

\section{Model}
\begin{figure}[bh]
  \begin{center}
     \includegraphics[width=0.85\linewidth]{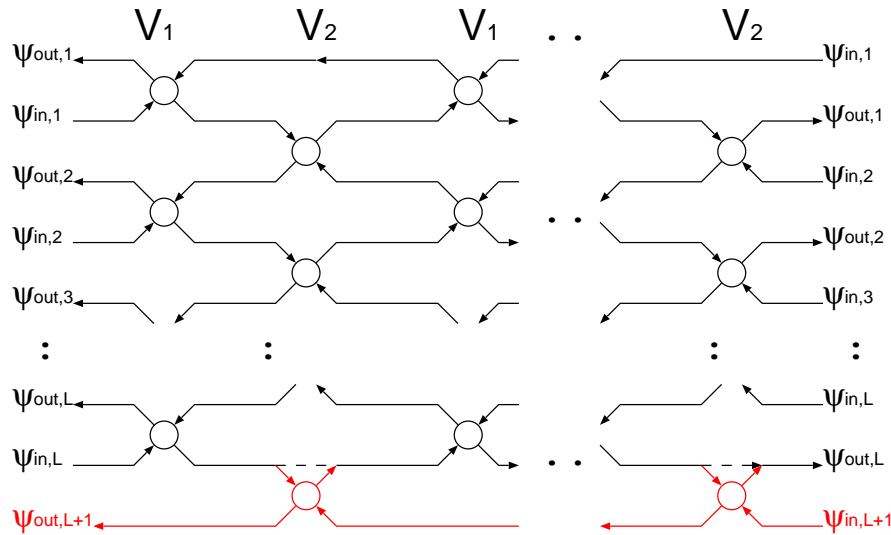}
  \end{center}
  \caption{Schematic of model. Note the lowest links that dramatically change the transport properties.}
  \label{NovelModel}
\end{figure}

From \textbf{S}$_2$, we define a 2 by 2 transfer matrix \textbf{T}$_2$
that relates the currents in the left side of the node to those in the right,
\begin{equation}
\textbf{T}_2=\left(
\begin{array}{cc}
1/t&r/t\\
r/t&1/t
\end{array}
\right)\,,
\end{equation}
and then construct the full transfer matrix \textbf{T}.
It consists of two types of transfer matrices.
One is the transfer matrix (see the left most column of  nodes in Fig. \ref{NovelModel}),
\begin{equation}
\textbf{V}_1=\left(
\begin{array}{ccccc}
\begin{array}{cc}
1/r & t/r \\
t/r & 1/r
\end{array} & 0_2 &\cdots & \begin{array}{c} 0 \\ 0\end{array}\\
0_2 & \ddots & \ddots & \vdots \\
\vdots  & \ddots &
\begin{array}{cc}
1/r & t/r \\
t/r & 1/r
\end{array} & \begin{array}{c} 0 \\ 0\end{array}\\
\begin{array}{ll} 0\,\, & \,\,0\end{array}
& \cdots &\begin{array}{ll} 0\,\, & \,\,0\end{array} & 1
\end{array}
\right)
\end{equation}
where $0_2$ is a $2\times 2$ zero matrix, and the other is
(see the next column in Fig. \ref{NovelModel})
\begin{equation}
\textbf{V}_2=\left(
\begin{array}{ccccc}
1 & \begin{array}{ll} 0\,\, & \,\,0\end{array} &\cdots & \begin{array}{ll} 0\, & \,0\end{array} \\
\begin{array}{cc} 0 \\ 0\end{array} &
\begin{array}{cc}
1/r & t/r \\
t/r & 1/r
\end{array} & 0_2 \,\, \cdots & 0_2\\
\vdots & 0_2 & \ddots & \vdots \\
\begin{array}{c} 0 \\ 0\end{array} & \cdots  & \ddots &
\begin{array}{cc}
1/r_\mathrm{ext} & t_\mathrm{ext}/r_\mathrm{ext} \\
t_\mathrm{ext}/r_\mathrm{ext} & 1/r_\mathrm{ext}
\end{array} \\
\end{array}
\right)\,.
\end{equation}
Here the new parameters $r_\mathrm{ext}$ and $t_\mathrm{ext}$
describe scattering due to the addition of a new link.
Between the nodes, we assume that the electron wavefunctions acquire
random phases $\mathrm{e}^{\mathrm{i}\phi}$, with
$\phi$ uniform random numbers on $[0,2\pi)$.
When the number of links in a layer is $N$, we need a set of $N$ independent
random number $\{ \phi \}$.
The transfer matrix $\textbf{T}$ that relates current fluxes in the left side
to that in the right side is
\begin{equation}
\textbf{T}=\prod_i V_\mathrm{D}(\{ \phi_i' \}) V_1 V_\mathrm{D}(\{ \phi_i \})V_2,
\end{equation}
where $i$ indicates $i$--th layer, and $V_\mathrm{D}(\{ \phi_i \})=
\mathrm{diag}[\mathrm{e}^{\mathrm{i}\phi_1},\cdots,\mathrm{e}^{\mathrm{i}\phi_N}]$.
The transfer matrix is then related to the reflection and transmission matrices
\cite{Kramer}.

To focus on the quantum Hall critical point, we set $r=t=1/\sqrt{2}$.
Unless explicitly stated, we assume $r=r_\mathrm{ext}$ and
$t=t_\mathrm{ext}$.

Before reporting our numerical results we describe the
structure of the  scattering matrix \textbf{S}.
This matrix relates the incoming and outgoing particle flux amplitudes
\begin{equation}
 \left(
  \begin{array}{@{\,}c@{\,}}
   \boldmath{\Psi}_\mathrm{L}^\mathrm{out} \\
   \boldmath{\Psi}_\mathrm{R}^\mathrm{out}
  \end{array}
 \right)=
 \textbf{S}
 \left(
  \begin{array}{@{\,}c@{\,}}
   \boldmath{\Psi}_\mathrm{L}^\mathrm{in} \\
   \boldmath{\Psi}_\mathrm{R}^\mathrm{in}
  \end{array}
 \right)=
 \left(
  \begin{array}{@{\,}cc@{\,}}
   \textbf{r} & \textbf{t}^{\prime} \\
   \textbf{t} & \textbf{r}^{\prime}
  \end{array}
 \right)
 \left(
  \begin{array}{@{\,}c@{\,}}
   \boldmath{\Psi}_\mathrm{L}^\mathrm{in} \\
   \boldmath{\Psi}_\mathrm{R}^\mathrm{in}
  \end{array}
 \right).
\end{equation}
Here $\boldmath{\Psi}_\mathrm{L(R)}^\mathrm{in(out)}$ is the incoming
(outgoing) current flux amplitude in the left (right) side of the system.
The $\textbf{t}$-matrix and $\textbf{t}^{\prime}$-matrix are $L\times L$
and $(L+1)\times (L+1)$ square matrices, respectively, while $\textbf{r}$
is $(L+1)\times L$ and $\textbf{r}^{\prime}$ is $L\times (L+1)$.
As a result, both $\textbf{S}$ and $\textbf{T}$ have odd dimensions $(2L+1)\times (2L+1)$.
The asymmetric C-C model is a type of quantum railroad \cite{Barnes,Barnes2}.

\section{Results}
The dimensionless conductances $G$ and $G^{\prime}$
\begin{equation}
  G=\mathrm{Tr}\textbf{t}\textbf{t}^{\dagger}\,\,\,,\,\,\,
  G^{\prime}=\mathrm{Tr}\textbf{t}^{\prime}\textbf{t}^{\prime\dagger},
\end{equation}
describe current from left to right and right to left, respectively.
Using the relationships obtained from the unitarity of the scattering matrix,
\begin{equation}
\textbf{t}^{\prime\dagger}\textbf{t}^{\prime}
+\textbf{r}^{\prime\dagger}\textbf{r}^{\prime}=\textbf{I}_{L+1} \,\,\,,\,\,\,
\textbf{t}\textbf{t}^{\dagger}
+\textbf{r}^{\prime}\textbf{r}^{\prime\dagger}=\textbf{I}_{L},
\end{equation}
and performing a singular value decomposition of $\textbf{r}^{\prime}$, we obtain
\begin{equation}
  \textbf{r}^{\prime}(L,L+1)
   =\textbf{U}(L,L)
     \left(
      \begin{array}{@{\,}ccccc@{\,}}
       \lambda_{1} & 0 & 0 & \cdots & 0 \\
       0 & \lambda_{2} & 0 & & \vdots \\
       \vdots & & \ddots & & \\
       0 & \cdots & & \lambda_{L} & 0 \\
      \end{array}
     \right)
    \textbf{U}'(L+1,L+1),
\end{equation}
where $\textbf{U}$ and $\textbf{U}'$ are unitary matrices and
$1\ge\lambda_{i}\ge 0$ ($i=1,2,\cdots, L$).
Taking its Hermitian conjugate,
\begin{equation}
  \textbf{r}^{\prime\dagger}(L+1,L)
   ={\textbf{U}'}^{\dagger}(L+1,L+1)
     \left(
      \begin{array}{@{\,}cccc@{\,}}
       \lambda_{1} & 0 & \cdots & 0 \\
       0 & \lambda_{2} & 0 & \vdots \\
       \vdots & & \ddots & \\
       & & & \lambda_{L} \\
       0 & \cdots & & 0
      \end{array}
     \right)
    \textbf{U}^{\dagger}(L,L),
\end{equation}
we obtain
\begin{equation}
  \textbf{t}\textbf{t}^{\dagger}
 =\textbf{U}\mathrm{diag}(1-\lambda_{1}^{2},1-\lambda_{2}^{2},
  \cdots,1-\lambda_{L}^{2})
  \textbf{U}^{\dagger}
\end{equation}
and
\begin{equation}
  \textbf{t}^{\prime\dagger}\textbf{t}^{\prime}
 ={\textbf{U}'}^{\dagger}\mathrm{diag}(1-\lambda_{1}^{2},1-\lambda_{2}^{2},
  \cdots,1-\lambda_{L}^{2},1)
  {\textbf{U}'}.
\end{equation}
Note that the transmission eigenvalues are the same except for a single extra unit
eigenvalue that appears in $\textbf{t}^{\prime\dagger}\textbf{t}^{\prime}$.
As a result, the difference between $G^{\prime}$ and $G$ is always unity,
\begin{equation}
   G^{\prime}-G=1.
\end{equation}
We might expect that the conductance distribution simply shifts by unity but,
as we show below, this expectation is too naive.

In the following, we calculate the transmission matrix $\bf{t}$ via the
transfer matrix technique.  The system width $L$ corresponds to the number
of nodes in a column, while the system length $M$ to that in a row.

\subsection{Conductance distribution for 2D}
\begin{figure}[bh]
  \begin{center}
     \includegraphics[width=0.85\linewidth]{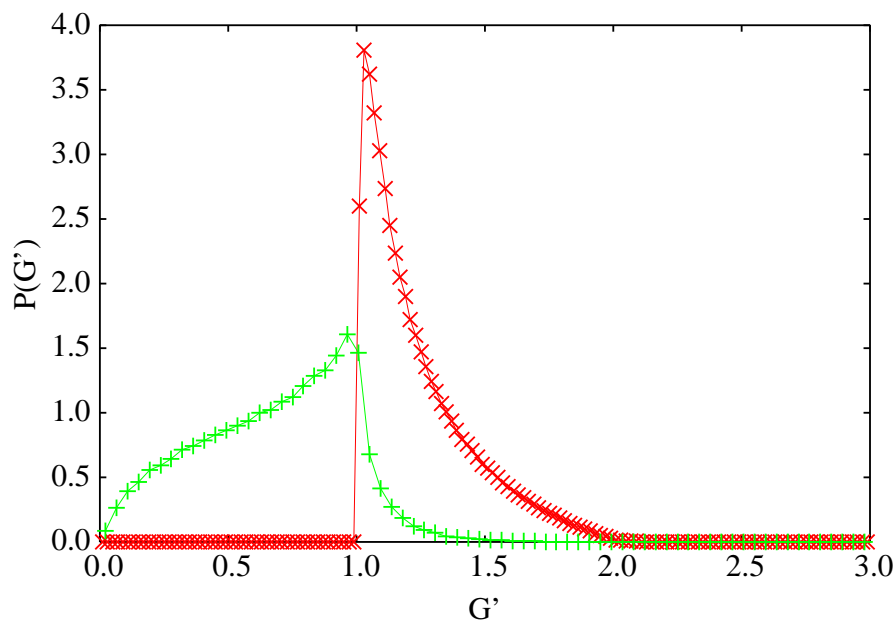}
  \end{center}
  \caption{Conductance distribution at critical point for the asymmetric CC model
  ($\times$).
  For comparison, results for the standard CC model (+) is plotted.
  The conductance is measured in units of $e^2/h$. $L=M=36$.  $10^6$ different
  random configurations have been realized.}
  \label{Distribution}
\end{figure}

The distribution of the conductance $G'$ in 2D
at the critical point is shown in Fig. \ref{Distribution}.
$L=M=32$ and $10^6$ samples have been realized.
The conductance is always greater than unity as expected.
The distribution has a kink at unity, and decreases rapidly with increasing $G'$.
This is explained by the fact that the unit eigenvalue repels the other eigenvalues.
This behavior is profoundly different from
that of the standard C-C model, where a broad distribution is found \cite{Takane}.

\subsection{Averaged conductances in 2D and quasi-1D}
\begin{figure}[bh]
  \begin{center}\leavevmode
     \includegraphics[width=0.85\linewidth]{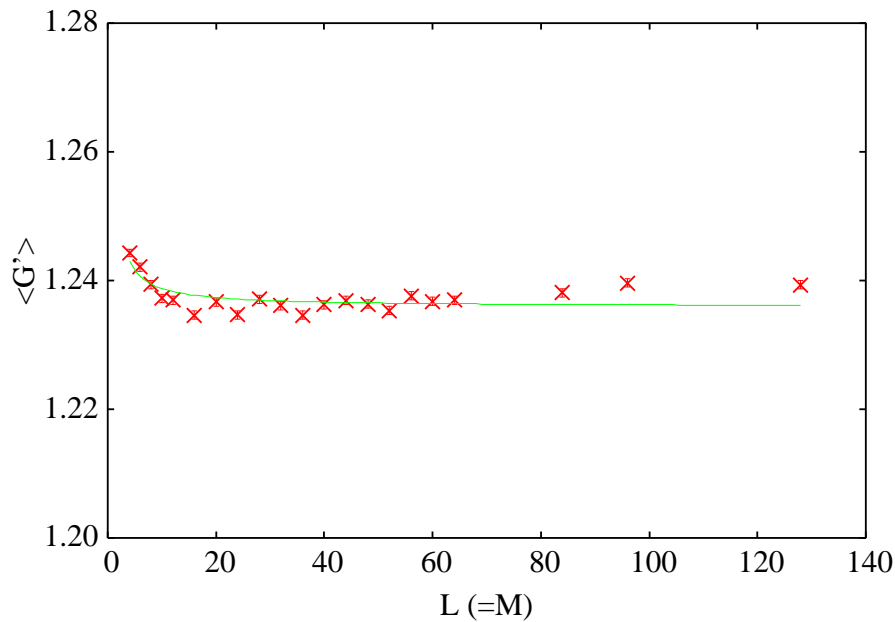}
    \end{center}
  \caption{Averaged conductance as function of system size $L$ for square geometry.
  Average over $10^6$ samples has been performed for each size.}
  \label{averagedCON}
\end{figure}

\begin{figure}[bh]
  \begin{center}
    \includegraphics[width=0.8\linewidth]{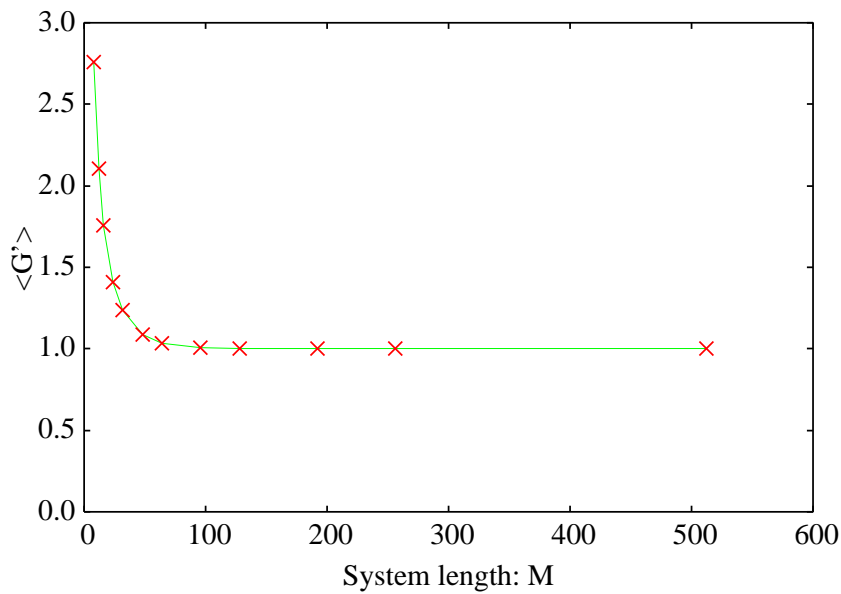} \\
   \end{center}
  \caption{Averaged conductance as function of system length $M$ in the quasi-1D wire.
  Average over $10^5$ samples has been performed.
  The width  $L=32$.}
  \label{averagedCON2}
\end{figure}

The averaged conductances $\langle G'\rangle$ for 2D and quasi-1D geometries are 
shown in Figs. \ref{averagedCON}
and  \ref{averagedCON2}.
In the limit $L\to\infty$, the conductance converges to
a finite value of, approximately, $1.24$ for the 2D geometry (Fig. \ref{averagedCON})
and to unity for the quasi-1D geometry (Fig. \ref{averagedCON2}).
We see that the asymmetric C-C model exhibits metallic behavior even for a long wire.
This is again very different to the standard
C-C model where the conductance converges to zero.
Also, the convergence to the limiting value is faster than that for the standard C-C model.
This is consistent with predictions based on random matrix theory
of an increase of the Lyapunov exponents associated with localised channels
due to the repulsion with the unit eigenvalue \cite{Takane}.

\section{Summary and Concluding Remarks}
In summary, we have proposed an asymmetric C-C model and analyzed its
transport properties.
We find that, even for a long wire, the conductance remains finite
and the system is metallic.
Transmission eigenvalues are repelled by the unit eigenvalue resulting
in an unusual form for the conductance distribution.

We have also varied the coupling to the additional link, $r_\mathrm{ext}$ and $t_\mathrm{ext}$.
Preliminary results suggest that conductance in sufficiently large systems is not sensitive to
the choice of coupling, indicating that
the observed transport phenomena are of bulk origin.

\subsection*{Acknowledgement}
We  acknowledge valuable discussions with T. Kawarabayashi,
Y. Takane, K. Wakabayashi and K. Kobayashi.
This work was supported by Grant-in-Aid No. 18540382 from MEXT.


\end{document}